\documentclass[aps,amsfonts,amsmath,prd,nofootinbib,tightenlines,superscriptaddress,10pt]{revtex4-2}

\usepackage{epsfig}
\usepackage{bm}
\usepackage{bbm}

\newcommand{\wavenumber}{\kappa}
\newcommand{\wavenumbersq}{\kappa^2}
\newcommand{\zint}{}
\newcommand{\wavevars}{\kappa}
\newcommand{\plusdsqdzsq}{}
\newcommand{\greenspherical}{}
\newcommand{\deltaproduct}{\delta(r-r') \delta(\theta-\theta')}
\newcommand{\plusdzsq}{}

\begin{document}

\title{Quantum Energy Density of Cosmic Strings with Nonzero Radius}


\author{Mao Koike}
\affiliation{Department of Physics, Middlebury College,
Middlebury, Vermont 05753, USA}

\author{Xabier Laquidain}
\affiliation{Department of Physics, Middlebury College,
Middlebury, Vermont 05753, USA}

\author{Noah Graham}
\email{ngraham@middlebury.edu}
\affiliation{Department of Physics, Middlebury College,
Middlebury, Vermont 05753, USA}

\begin{abstract}
Zero-point fluctuations in the background of a cosmic string provide an
opportunity to study the effects of topology in quantum field theory.
We use a scattering theory approach to compute quantum corrections to
the energy density of a cosmic string, using the ``ballpoint pen'' and
``flowerpot'' models to allow for a nonzero string radius.  For
computational efficiency, we consider a massless field in $2+1$
dimensions.  We show how to implement precise and unambiguous
renormalization conditions in the presence of a deficit angle, and
make use of Kontorovich-Lebedev techniques to rewrite the sum over
angular momentum channels as an integral on the imaginary axis. 
\end{abstract}

\maketitle

\section{Introduction}

The effects of quantum fluctuations can be of particular importance in
systems with nontrivial topology.  In one space
dimension, one can carry out calculations analytically in scalar models
\cite{Dashen:1974ci}, and their supersymmetric extensions.  In the
latter case, such corrections appear to violate the 
Bogomol'nyi-Prasad-Sommerfield bound
\cite{Rebhan:1997iv}, but a corresponding correction to the central
charge ensures that the bound remains saturated
\cite{super1d,Dunne:1999du,SVV}.  In higher dimensions, such
corrections must vanish identically in supersymmetric models to
preserve multiplet shortening \cite{Witten:1978mh}, while detailed
calculations are difficult in nonsupersymmetric models.  String
backgrounds in Higgs-gauge theory offer the opportunity to study
quantum effects of topology in higher dimensions while remaining
computationally tractable, making it possible to study quantum effects
on string stability \cite{Weigel:2010zk,PhysRevD.104.L011901}.

In this paper we consider quantum fluctuations of a massless scalar
field $\phi$ in the gravitational background of a cosmic string, which
introduces topological effects through a deficit angle in the
otherwise flat spacetime outside the string core.  When the string is
taken to have zero radius, the problem becomes scale invariant and the
quantum corrections can be computed exactly
\cite{PhysRevD.34.1918,PhysRevD.35.536,PhysRevD.35.3779}.
For a string of nonzero radius $r_0$, we must specify a profile
function for the background curvature, which was previously
concentrated at $r=0$.  We will consider two such models
for the string core
\cite{PhysRevD.31.3288,1985ApJ...288..422G,PhysRevD.42.2669}, the
``ballpoint pen,'' in which the curvature is
constant for $r<r_0$, and the ``flowerpot,'' in which the curvature is
localized to a $\delta$-function ring at $r=r_0$.  In the former case,
the curvature for a specified deficit angle is chosen so that the
metric is continuous at $r_0$.  For both models, we can construct the
scattering wavefunctions \cite{PhysRevD.42.2669}, matching the
solutions inside and outside using boundary conditions at $r=r_0$.
From these scattering data, we can then construct the Green's
function, from which we determine the quantum energy density.  This
calculation is formally divergent and requires renormalization.  In
all cases, we must subtract the contribution of the free Green's
function, which corresponds to renormalization of the cosmological
constant.  Because of the nontrivial topology, this subtraction is
most efficiently implemented by adding and subtracting the result for
the zero radius ``point string,'' and then using analytic continuation
to imaginary angular momentum to compute the difference between the
point string and the free background.  In addition, for the interior of
the ballpoint pen, we have a nontrivial background potential and must
subtract the tadpole contribution, which corresponds to a
renormalization of the gravitational constant.

In this calculation, we find it advantageous to break the calculation of the
quantum energy density into two parts:  a ``bulk'' term $\langle
(\partial_t \phi)^2 \rangle$ and a ``derivative'' term
$\displaystyle \left(\frac{1}{4}- \xi \right) \left\langle 
\frac{1}{r^2} {\cal D}_r^2 (\phi^2) \right\rangle$, where $\xi$ is the
curvature coupling, for which we can identify the counterterm
contributions individually.  Putting these results together, we obtain
the full quantum energy density as a function of $r$ for a given
deficit angle, string radius, and curvature coupling, which we can
efficiently compute as a numerical sum and integral over the
fluctuation spectrum.

\section{Model and Green's Function}

We begin from the general case of scalar field in $d$ spacetime
dimensions, for which the action functional is
\begin{equation}
S = -\frac{1}{2}\int d^d x \sqrt{-g} \left(
\nabla_\alpha \phi \nabla^\alpha \phi + U \phi^2 
+ \xi {\cal R} \phi^2\right) \,,
\end{equation}
with coupling $\xi$ to the Ricci curvature scalar ${\cal R}$.  Of
particular interest is the case of conformal coupling, $\displaystyle
\xi = \frac{1}{8}$ in two space dimensions and $\displaystyle \xi =
\frac{1}{6}$ in three space dimensions.  This expression includes
an external background potential $U$; we will set $U=0$, but it is
straightforward to introduce a nontrivial $U(\phi)$, such
as a mass term $U=\mu^2$.  The equation of motion is
\begin{equation}
-\nabla_\alpha\nabla^\alpha\phi + U\phi + \xi {\cal R} \phi = 0
\end{equation}
with metric signature $(-+++)$.  The stress-energy tensor is given by
\cite{PhysRevD.54.6233,PhysRevD.72.065013,fliss2023}
\begin{equation}
T_{\alpha\beta} = \nabla_\alpha \phi \nabla_\beta \phi
-g_{\alpha \beta} \frac{1}{2}  \left(\nabla_\gamma \phi 
\nabla^\gamma \phi + U \phi^2\right)
+ \xi \phi^2 \left( R_{\alpha\beta} -\frac{1}{2} g_{\alpha \beta} {\cal R}
\right)
+ \xi \left(g_{\alpha \beta} \nabla_\gamma \nabla^\gamma
- \nabla_\alpha \nabla_\beta\right)(\phi^2)\,,
\label{eqn:stressenergy}
\end{equation}
as obtained by varying the action with respect to the metric.  Note
that the curvature coupling contributes to the stress-energy tensor even
in regions where ${\cal R}=0$, although it does so by a total derivative.

We consider the spacetime metric
\begin{equation}
ds^2 = -dt^2 \plusdsqdzsq + p(r)^2 dr^2 + r^2 d\theta^2
\end{equation}
with a deficit angle $2\theta_0$, meaning that the
range of angular coordinate is $0\ldots 2(\pi-\theta_0)$, and we
define $\displaystyle \sigma = \frac{\pi}{\pi-\theta_0}$.  To
implement the deficit angle without a singularity at the origin, we
introduce a profile function $p(r)$ that ranges from 
$\displaystyle \frac{1}{\sigma}$ at the origin to $1$ at the string
radius $r_0$.  The nonzero Christoffel symbols in this geometry are
\begin{equation}
\Gamma_{r r}^r =  \frac{p'(r)}{p(r)} \qquad
\Gamma_{\theta \theta}^r = -\frac{r}{p(r)^2} \qquad
\Gamma_{\theta r}^\theta = \Gamma_{r \theta}^\theta = \frac{1}{r}\,,
\end{equation}
and because the geometry only has curvature in two dimensions, all the
nonzero components of the Riemann and Ricci tensors
\begin{equation}
R_{\theta r\theta}^r = -R_{\theta\theta r}^r = 
R_{\theta\theta} = g_{\theta\theta} \frac{{\cal R}}{2}
\qquad
R_{r\theta r}^\theta = -R_{rr\theta}^\theta
=  R_{rr} = g_{rr} \frac{{\cal R}}{2}
\end{equation}
can be expressed in terms of the curvature scalar $\displaystyle
{\cal R}=\frac{2}{r} \frac{p'(r)}{p(r)^3}$.  Acting on any scalar
$\chi$, the covariant derivatives simply become ordinary derivatives,
while for second derivatives we have nontrivial contributions from the
Christoffel symbols given above,
\begin{equation}
\nabla_\theta \nabla_\theta \chi = \partial_\theta^2 \chi -
\Gamma_{\theta\theta}^r \partial_r \chi \qquad
\nabla_r \nabla_r \chi = \partial_r^2 \chi -
\Gamma_{rr}^r \partial_r \chi \qquad
\nabla_r \nabla_\theta \chi =
\nabla_\theta \nabla_r \chi = \partial_\theta \partial_r \chi -
\Gamma_{r\theta}^\theta \partial_\theta \chi \,,
\end{equation}
and, as a result, covariant derivatives with respect to $\theta$
can be nonzero even if $\chi$ is rotationally invariant.  In
particular, we have
\begin{equation}
(g^{\theta \theta} \nabla_\theta \nabla_\theta + g^{rr} \nabla_r
\nabla_r) \chi = \frac{1}{r^2} \left(\frac{\partial^2 \chi}
{\partial \theta^2} + {\cal D}_r^2 \right)\chi,
\end{equation}
where $\displaystyle {\cal D}_r = \frac{r}{p(r)} \frac{\partial}{\partial r}$
is the radial derivative.  

\begin{figure}[htbp]
\includegraphics[width=0.5\linewidth]{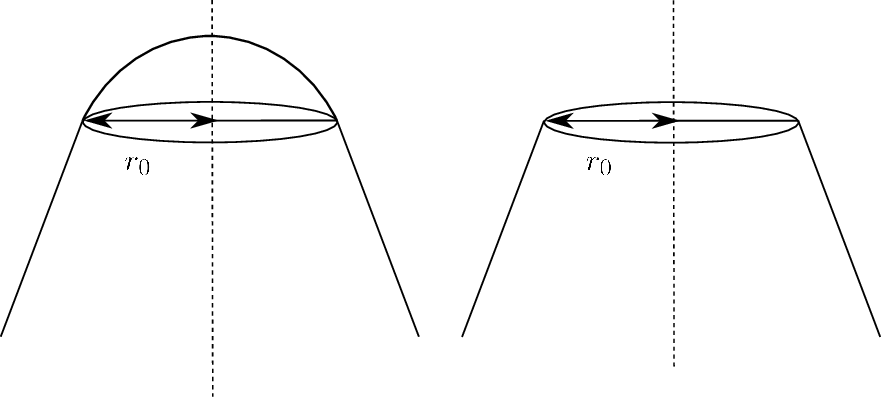}
\caption{Illustration of the ballpoint pen (left) and flowerpot
(right) string geometries.}
\label{fig:models}
\end{figure}

By the symmetry of the string configuration and vacuum state, the
vacuum expectation value of $\phi^2$ will depend only on $r$, and we
can write the energy density as
\begin{equation}
\langle T_{tt} \rangle =\left\langle
\frac{1}{2} (\partial_t \phi)^2 +\frac{1}{2r^2}({\cal D}_r \phi)^2
+ \frac{1}{2r^2}(\partial_\theta \phi)^2 \plusdzsq
+ \frac{\xi}{2} {\cal R}\phi^2 - \frac{\xi}{r^2} {\cal D}_r^2
(\phi^2) \right\rangle \,,
\end{equation}
where $\phi$ obeys the equation of motion
\begin{equation}
\left(\frac{\partial^2}{\partial t^2} - \frac{1}{r^2} {\cal D}_r^2
 - \frac{1}{r^2} \frac{\partial^2}{\partial \theta^2} 
\plusdsqdzsq + \xi {\cal R} \right) \phi = 0 \,.
\end{equation}
Again by symmetry, we have $\displaystyle \frac{1}{2} \partial_t^2 
\langle \phi^2 \rangle = \partial_t \langle \phi (\partial_t \phi)
\rangle=0$, and so $\langle \phi (\partial_t^2 \phi) \rangle = -
\langle (\partial_t \phi)^2 \rangle$, and similarly for
$\theta$. 
Using these results and $\displaystyle {\cal D}_r^2 (\phi^2) = 
2 ({\cal D}_r \phi)^2 + 2 \phi {\cal D}_r^2 \phi$, together with the
equation of motion, we obtain
\begin{equation}
\left\langle \frac{1}{4 r^2} {\cal D}_r^2 (\phi^2) \right\rangle =
\left\langle \frac{1}{2r^2}({\cal D}_r \phi)^2 
-\frac{1}{2} (\partial_t \phi)^2 + \frac{1}{2r^2}
(\partial_\theta \phi)^2 \plusdzsq + \frac{\xi}{2} {\cal R}\phi^2
\right\rangle\,,
\end{equation}
yielding a simplified expression in terms of a consolidated derivative term,
\begin{equation}
\langle T_{tt} \rangle = \left\langle
(\partial_t \phi)^2 + \left(\frac{1}{4}- \xi \right) 
\frac{1}{r^2} {\cal D}_r^2 (\phi^2) \right\rangle \,.
\end{equation}
This form will be more convenient for organizing the calculation, 
particularly with regard to renormalization using the techniques of
Ref.\ \cite{OLUM2003175}, but for numerical calculation we will find
it preferable to re-expand the second derivative in terms of squared
first derivatives.

Our primary tool will be the Green's function
$G_\sigma(\bm{r},\bm{r}',\kappa)$ for imaginary wave number
$k=i\kappa$, which obeys
\begin{equation}
-\left(\frac{1}{r^2} {\cal D}_r^2
+ \frac{1}{r^2} \frac{\partial^2}{\partial \theta^2}
\plusdsqdzsq - \xi {\cal R} - \wavenumbersq \right)
G_\sigma(\bm{r},\bm{r}',\kappa)  = \frac{1}{r p(r)} \deltaproduct \,.
\end{equation}
We consider two profile functions 
\cite{PhysRevD.31.3288,1985ApJ...288..422G,PhysRevD.42.2669}: 
the flowerpot,
\begin{equation}
p^{\rm flower}(r) = 
\begin{cases}
\frac{1}{\sigma} & r< r_0 \\
1 &r>r_0\end{cases} \,,
\end{equation}
and the ballpoint pen, 
\begin{equation}
p^{\rm pen}(r) = 
\begin{cases}
\left[\sigma^2-\frac{r^2}{r_0^2} (\sigma^2-1)\right]^{-1/2}&r<r_0\\
1 &r>r_0\end{cases}\,,
\end{equation}
where $r_0$ is the string radius, as shown in Fig.\ \ref{fig:models}.
The flowerpot has zero curvature everywhere except for a
$\delta$-function contribution at the string radius,  while the
ballpoint pen has constant curvature
$\displaystyle {\cal R}=\frac{2 (\sigma^2-1)}{r_0^2}$
inside and zero curvature outside.  We write the Green's function in
the scattering form
\begin{equation}
G_\sigma(\bm{r},\bm{r}',\kappa) = \frac{\sigma}{\pi}
\zint
\sum_{\ell=0}^\infty{}'
\psi_{\wavevars,\ell}^{\rm reg} (r_<) \psi_{\wavevars,\ell}^{\rm out} (r_>)
\cos \left[\sigma \ell (\theta-\theta')\right] \,,
\end{equation}
where the prime on the sum indicates that the $\ell=0$ term is counted
with a weight of one-half, arising because we have written the sum
over nonnegative $\ell$ only.  The radial wavefunctions obey the equation
\begin{equation}
\left[-\frac{1}{r^2} {\cal D}_r^2 + 
\frac{\ell^2\sigma^2}{r^2} + \xi {\cal R} + \wavenumbersq
\right]\psi_{\wavevars,\ell}(r) = 0 \,,
\label{eqn:eom}
\end{equation}
where the regular solution is
defined to be well-behaved at $r=0$, while the outgoing solution
obeys outgoing wave boundary conditions for $r\to \infty$, normalized
to unit amplitude, and $r_<$ ($r_>$) is the smaller (larger) axial
radius of $\bm{r}$ and $\bm{r}'$. The functions are normalized so that
they obey the Wronskian relation
\begin{equation}
\frac{d}{dr} \left(\psi_{\wavevars,\ell}^{\rm reg}(r)\right)
\psi_{\wavevars,\ell}^{\rm out}(r)
- \psi_{\wavevars,\ell}^{\rm reg}(r) \frac{d}{dr} 
\left(\psi_{\wavevars,\ell}^{\rm out}(r)\right)
= \frac{p(r)}{r} \,,
\label{eqn:wronk}
\end{equation}
which provides the appropriate jump condition for the Green's function.

As shown in Ref.\ \cite{OLUM2003175}, the renormalized energy density
of a scalar field in flat spacetime with a background potential
that is spherically symmetric in $m$ dimensions and independent of $n$
dimensions after a single ``tadpole'' subtraction can be written as
\begin{eqnarray}
\langle{\cal H}\rangle_{\rm ren} 
&=& -\frac{1}{2(4\pi)^{\frac{n+1}{2}}\Gamma\left(\frac{n+3}{2}\right)}
\sum_{\ell=0}^\infty D^m_\ell \int_0^\infty d\kappa \, 2\kappa^{n+2} \left[  
{\cal G}_{\ell,m}(r,r, \kappa) - {\cal G}_{\ell,m}^{\rm free}(r,r, \kappa)
\phantom{\frac{1}{1}}\right. \cr && \left. \times
\left(1 - (2-m)\frac{V_\ell(r)}{2\kappa^2}\right)
-\frac{n+1}{\kappa^2}
\left(\frac{1}{4} - \xi\right)\frac{1}{r^{2m-2}}
{\cal D}^2_{r,m} {\cal G}_{\ell,m}(r,r, \kappa) \right] \,,
\label{eqn:onesub}
\end{eqnarray}
where we have included the contribution from the curvature coupling
$\xi$, which contributes to Eq.\ (\ref{eqn:stressenergy}) even when
${\cal R}=0$.  Here $\displaystyle \frac{1}{r^{2m-2}}
{\cal D}^2_{r,m}$ with $\displaystyle {\cal D}_{r,m} =
\frac{r^{m-1}}{p(r)} \frac{\partial}{\partial r}$
is the radial Laplacian in $m$ dimensions,
$V_\ell(r)$ is the scattering potential in channel $\ell$
with degeneracy factor $D^m_\ell$, and we have decomposed the
$m$-dimensional Green's function for equal angles into its component
${\cal G}_{\ell,m}(r,r',\kappa)$ in each channel, which obeys the equation
\begin{equation}
\left(-{\cal D}_{r,m}^2 + V_\ell(r) +
\frac{\ell(\ell+m-2)}{r^2} + \kappa^2\right)
{\cal G}_{\ell,m}(r,r',\kappa) = \delta^{(m)}(r-r')
\end{equation}
in terms of the $m$-dimensional $\delta$-function.
We will focus on the case of $m=2$ and $n=0$, so a single subtraction
will be sufficient.  The case of a three-dimensional string with $m=2$
and $n=1$ works similarly, but requires additional renormalization
counterterms due to the higher degree of divergence.

\section{Point String and Kontorovich-Lebedev Approach}
\label{sec:pointstring}

We begin by reviewing the case of the point string, 
\cite{PhysRevD.34.1918,PhysRevD.35.536,PhysRevD.35.3779}
where the radius $r_0$ of the string core is taken to zero.  The
scattering solutions can be obtained using the same techniques as for
a conducting wedge \cite{PhysRevD.20.3063,BREVIK1996157}, but with
periodic rather than perfectly reflecting boundary conditions.  The
normalized scattering functions are
$\psi_{\wavevars,\ell}^{\rm reg,point}(r) = 
I_{\sigma \ell}\left(\wavenumber r\right)$ and
$\psi_{\wavevars,\ell}^{\rm out,point}(r) = 
K_{\sigma \ell}\left(\wavenumber r\right)$,
and the Green's function becomes
\begin{equation}
G_\sigma^{\rm point}(\bm{r},\bm{r}',\kappa) = \frac{\sigma}{\pi}
\zint
\sum_{\ell=0}^\infty{}' I_{\sigma \ell}\left(\wavenumber r_<\right)
K_{\sigma \ell}\left(\wavenumber r_>\right)
\cos \left[\ell \sigma (\theta-\theta')\right] \,,
\label{eqn:pointstring}
\end{equation}
Setting $\sigma=1$, we obtain the free Green's function
\begin{equation}
G^{\rm free}(\bm{r},\bm{r}',\kappa) = \frac{1}{\pi}
\zint
\sum_{\ell=0}^\infty{}' I_{\ell}\left(\wavenumber r_<\right)
K_{\ell}\left(\wavenumber r_>\right)
\cos \left[\ell (\theta-\theta')\right] 
= \frac{1}{2\pi} \zint K_0\left(\wavenumber\left|r
e^{i\theta}-r'e^{i\theta'}\right|\right) \greenspherical \,.
\label{eqn:free}
\end{equation}

A useful computational tool is to replace the sum over
the angular quantum number $\ell$ by a contour integral,
based on the Kontorovich-Lebedev transformation.  In this approach
\cite{Oberhettinger,Maghrebi6867,physics5040065}, one multiplies the summand by
$\displaystyle \frac{\pi (-1)^\ell}{\sin \pi \ell}$, which has poles
of unit residue at all integers $\ell$.  Because the summand has
no other poles in the right half of the complex plane, the original
sum over nonnegative $\ell$ then equals the integral of this product 
over a contour that goes down the imaginary axis and returns by a
large semicircle at infinity, taking into account the factor of $2\pi
i$ from Cauchy's theorem.  The infinitesimal semicircle needed to go
around the pole at $\ell=0$ accounts for the factor of one-half
associated with that term in the sum. For the functions we consider,
the integral over the large semicircle vanishes, while the
contributions from the negative and positive imaginary axis can be
folded into a single integral, which often can be simplified through
identities such as
\begin{equation}
K_\nu(x)=\frac{\pi}{2}\frac{I_{-\nu}(x)-I_{\nu}(x)}{\sin{\nu \pi}}\,,
\end{equation}
which is valid for any $\nu$ that is not a real integer.

We illustrate this approach using the point string.  To compute finite
quantum corrections, we will want to take the difference between the
full Green's function in Eq.\ (\ref{eqn:pointstring}) and the free
Green's function in Eq.\ (\ref{eqn:free}), in the limit where the
points become coincident, meaning that the individual Green's functions
diverge. However, the necessary cancellation does not emerge
term-by-term in the sum, and as a result the standard calculations for
this case
\cite{PhysRevD.34.1918,PhysRevD.42.2669,BREVIK1996157} first carry out
the integral over $\kappa$, taking advantage of the availability of
analytic results in three space dimensions, which do not exist in our
case.

In contrast, using the Kontorovich-Lebedev approach as described
above, we obtain
\begin{equation}
G^{\rm free}(\bm{r},\bm{r}',\kappa) = \frac{1}{\pi^2}
\zint
\int_0^\infty d\lambda K_{i\lambda}(\wavenumber r)
K_{i\lambda}(\wavenumber r')
\cosh[\lambda(\pi - |\theta-\theta'|)]\,,
\label{eqn:KLfree}
\end{equation}
where we have used $\ell = i \lambda$ and the $\pi$ term in the
hyperbolic cosine reflects the $(-1)^\ell$ factor above.  Note that
the jump condition now emerges from the angular rather than the radial
component. Similarly, by letting $\sigma \ell = i \lambda$, we obtain
for the point string
\begin{equation}
G_\sigma^{\rm point}(\bm{r},\bm{r}',\kappa) = \frac{1}{\pi^2}
\zint
\int_0^\infty d\lambda K_{i\lambda}(\wavenumber r)
K_{i\lambda}(\wavenumber r')
\cosh\left[\lambda \left(\frac{\pi}{\sigma} - |\theta-\theta'|\right)\right]
\frac{\sinh \lambda \pi}{\sinh \frac{\lambda}{\sigma}\pi}\,,
\end{equation}
and the difference between Green's functions can be computed by
subtraction under the integral sign, yielding after simplification
\begin{equation}
\Delta G_\sigma^{\rm point}(r,r,\kappa)
=G_\sigma^{\rm point}(r,r,\kappa) - G^{\rm free}(r,r,\kappa) 
= \frac{1}{\pi^2} \zint
\int_0^\infty d\lambda K_{i\lambda}(\wavenumber r)
K_{i\lambda}(\wavenumber r)
\frac{\sinh \left[\frac{\lambda}{\sigma}
\pi (\sigma-1)\right]} {\sinh \frac{\lambda}{\sigma}\pi} \,,
\label{eqn:deltapointstring}
\end{equation}
where we have now taken the limit of coincident points since the
difference of Green's functions is nonsingular.

\section{Scattering Wavefunctions}

Following Ref.\ \cite{PhysRevD.42.2669},
we next compute the regular and outgoing scattering wavefunctions for
both the flowerpot and ballpoint pen, each of which will be computed
piecewise, with separate expressions inside and outside of the
string.  In regions where $p(r)$ is constant, namely for $r>r_0$ in
both models and $r<r_0$ for the flowerpot, we have ${\cal R}=0$ and the
solutions to Eq.\ (\ref{eqn:eom}) are modified Bessel functions 
$I_{\tilde \sigma \ell}(\wavenumber r_*)$ and
$K_{\tilde \sigma \ell}(\wavenumber r_*)$, where $\ell$ is an
integer, $\tilde \sigma = \sigma p(r)$, and $r_* = p(r) r$ is the
physical radial distance.  For $r<r_0$ in the ballpoint pen model, the
solutions are Legendre functions 
$P_{\nu(\wavevars)}^\ell\left(\tfrac{1}{\sigma  p(r)}\right)$
and $Q_{\nu(\wavevars)}^\ell\left(\tfrac{1}{\sigma  p(r)}\right)$
with $\nu(\wavevars) (\nu(\wavevars) + 1) = 
-\left(\frac{r_0^2\kappa^2}{\sigma^2 -1}+ 2 \xi\right)$,
so that
\begin{equation}
\nu(\wavevars) = -\frac{1}{2} + \frac{1}{2}\sqrt{(1-8\xi) - 
\frac{4\wavenumbersq r_0^2}{\sigma^2-1}} \,.
\end{equation}
We can thus write the full solutions as
\begin{equation}
\psi_{\wavevars,\ell}^{\rm pen}(r) =
\begin{array}{|c|c|c|}
\hline
& r<r_0 & r>r_0\cr
\hline
\hbox{regular} & A_{\wavevars,\ell}^{\rm pen}
P_{\nu(\wavevars)}^\ell\left(\tfrac{1}{\sigma  p(r)}\right) & 
I_{\sigma \ell}(\wavenumber r) + B_{\wavevars,\ell}^{\rm pen}
K_{\sigma \ell}(\wavenumber r)
\cr
\hline
\hbox{outgoing} &  
C_{\wavevars,\ell}^{\rm pen}
P_{\nu(\wavevars)}^\ell\left(\tfrac{1}{\sigma p(r)}\right) +
D_{\wavevars,\ell}^{\rm pen} 
Q_{\nu(\wavevars)}^\ell\left(\tfrac{1}{\sigma p(r)}\right)
& K_{\sigma \ell}(\wavenumber r) \cr
\hline
\end{array}
\end{equation}
for the ballpoint pen and
\begin{equation}
\psi_{\wavevars,\ell}^{\rm flower}(r) =
\begin{array}{|c|c|c|}
\hline
& r<r_0 & r>r_0\cr
\hline
\hbox{regular} & A_{\wavevars,\ell}^{\rm flower}
I_{\ell}\left(\wavenumber \frac{r}{\sigma}\right) &
I_{\sigma \ell}(\wavenumber r) + B_{\wavevars,\ell}^{\rm flower}
K_{\sigma \ell}(\wavenumber r)
\cr
\hline
\hbox{outgoing} &  
C_{\wavevars,\ell}^{\rm flower}
I_{\ell}\left(\wavenumber \frac{r}{\sigma}\right) +
D_{\wavevars,\ell}^{\rm flower} 
K_{\ell}\left(\wavenumber \frac{r}{\sigma}\right)
& K_{\sigma \ell}(\wavenumber r) \cr
\hline
\end{array}
\end{equation}
for the flowerpot.  In these expressions, for $r>r_0$ the coefficient
of the outgoing wave is normalized to one, and then we can also set the
coefficient of the first-kind solution in the regular wave to one by
the Wronskian relation, Eq.\ (\ref{eqn:wronk}).  For $r<r_0$, the
regular solution must be proportional to the first-kind function,
since it is the only solution regular at the origin.  In the
ballpoint pen model, both the wavefunction and its first derivative are
continuous at $r=r_0$, while in the flowerpot model the wavefunction and
the quantity
\begin{equation}
\frac{r}{p(r)} \frac{d}{dr} \psi_{\wavevars,\ell}^{\rm flower}(r) + 
2 \xi \frac{\psi_{\wavevars,\ell}^{\rm flower}(r)}{p(r)}
\end{equation}
are continuous at $r=r_0$ (note that $p(r)$ is discontinuous).  
The boundary conditions for the function and its first
derivative at $r=r_0$ thus yield four equations for the four unknown
coefficients.  In addition, from the Wronskian relation for $r<r_0$ we
know that
\begin{equation}
A_{\wavevars,\ell}^{\rm pen} D_{\wavevars,\ell}^{\rm pen} = 
\frac{1}{\sigma}\frac{\Gamma(\nu(\wavevars)-\ell +1)}
{\Gamma(\nu(\wavevars)+\ell +1)}
\hbox{\qquad and \qquad}
A_{\wavevars,\ell}^{\rm flower} D_{\wavevars,\ell}^{\rm flower} = 
\frac{1}{\sigma} \,.
\label{eqn:insidewronk}
\end{equation}

Given this result, for brevity we quote only the remaining
combinations we will need to form the Green's function,
\begin{eqnarray}
\frac{C_{\wavevars,\ell}^{\rm pen}}{D_{\wavevars,\ell}^{\rm pen}} &=& -
\frac{
(\sigma^2-1)Q_{\nu(\wavevars)}^\ell{}'\left(\tfrac{1}{\sigma}\right)
K_{\sigma \ell}\left(\wavenumber r_0\right) +
\sigma \kappa r_0 Q_{\nu(\wavevars)}^\ell\left(\tfrac{1}{\sigma}\right)
K_{\sigma \ell}'\left(\wavenumber r_0\right)
}{
(\sigma^2-1)P_{\nu(\wavevars)}^\ell{}'\left(\tfrac{1}{\sigma}\right)
K_{\sigma \ell}\left(\wavenumber r_0\right) +
\sigma \kappa r_0 P_{\nu(\wavevars)}^\ell\left(\tfrac{1}{\sigma}\right)
K_{\sigma \ell}'\left(\wavenumber r_0\right)
}
\cr
B_{\wavevars,\ell}^{\rm pen} &=& -\frac{
(\sigma^2-1)P_{\nu(\wavevars)}^\ell{}'\left(\tfrac{1}{\sigma}\right)
I_{\sigma \ell}\left(\wavenumber r_0\right) +
\sigma \kappa r_0 P_{\nu(\wavevars)}^\ell\left(\tfrac{1}{\sigma}\right)
I_{\sigma \ell}'\left(\wavenumber r_0\right)
}{
(\sigma^2-1)P_{\nu(\wavevars)}^\ell{}'\left(\tfrac{1}{\sigma}\right)
K_{\sigma \ell}\left(\wavenumber r_0\right) +
\sigma \kappa r_0 P_{\nu(\wavevars)}^\ell\left(\tfrac{1}{\sigma}\right)
K_{\sigma \ell}'\left(\wavenumber r_0\right)
}
\end{eqnarray}
and
\begin{eqnarray}
\frac{C_{\wavevars,\ell}^{\rm flower}}
{D_{\wavevars,\ell}^{\rm flower}} &=& -\frac{
K_{\ell}'\left(\wavenumber \frac{r_0}{\sigma}\right)
K_{\sigma \ell}\left(\wavenumber r_0\right)
-
K_{\ell}\left(\wavenumber \frac{r_0}{\sigma}\right)
K_{\sigma \ell}'\left(\wavenumber r_0\right)
+ \frac{2 \xi(\sigma-1)}{\wavenumber r_0}
K_{\ell}\left(\wavenumber \frac{r_0}{\sigma}\right)
K_{\sigma \ell}\left(\wavenumber r_0\right)
}{
I_{\ell}'\left(\wavenumber \frac{r_0}{\sigma}\right)
K_{\sigma \ell}\left(\wavenumber r_0\right)
- I_{\ell}\left(\wavenumber \frac{r_0}{\sigma}\right)
K_{\sigma \ell}'\left(\wavenumber r_0\right)
+ \frac{2 \xi(\sigma-1)}{\wavenumber r_0}  
I_{\ell}\left(\wavenumber \frac{r_0}{\sigma}\right)
K_{\sigma \ell}\left(\wavenumber r_0\right)
}
\cr
B_{\wavevars,\ell}^{\rm flower} &=& -\frac{
I_{\ell}'\left(\wavenumber \frac{r_0}{\sigma}\right)
I_{\sigma \ell}\left(\wavenumber r_0\right) -
I_{\ell}\left(\wavenumber \frac{r_0}{\sigma}\right)
I_{\sigma \ell}'\left(\wavenumber r_0\right)
+ \frac{2 \xi(\sigma-1)}{\wavenumber r_0}
I_{\ell}\left(\wavenumber \frac{r_0}{\sigma}\right)
I_{\sigma \ell}\left(\wavenumber r_0\right)
}{
I_{\ell}'\left(\wavenumber \frac{r_0}{\sigma}\right)
K_{\sigma \ell}\left(\wavenumber r_0\right) 
- I_{\ell}\left(\wavenumber \frac{r_0}{\sigma}\right)
K_{\sigma \ell}'\left(\wavenumber r_0\right)
+ \frac{2 \xi(\sigma-1)}{\wavenumber r_0}
I_{\ell}\left(\wavenumber \frac{r_0}{\sigma}\right)
K_{\sigma \ell}\left(\wavenumber r_0\right)
} \,,
\end{eqnarray}
where prime denotes a derivative with respect to the function's argument.

Finally, we note that when $\ell$ is not a real integer, as will arise
in situations we consider below, for $r<r_0$
it is computationally preferable to take as independent solutions
$P_{\nu(\wavevars)}^\ell$ and $P_{\nu(\wavevars)}^{-\ell}$  
rather than 
$P_{\nu(\wavevars)}^\ell$ and $Q_{\nu(\wavevars)}^{\ell}$ 
for the ballpoint pen, and
$I_{\ell}$ and $I_{-\ell}$ 
rather than
$I_{\ell}$ and $K_{\ell}$ 
for the flowerpot.  With these replacements made
throughout, the same formulae hold as above, except that the
right-hand side of Eq.\ (\ref{eqn:insidewronk}) becomes
$\displaystyle \frac{\pi}{2 \sigma \sin \pi \ell}$ in both cases.

\section{Renormalization: Free Green's Function Subtraction}
\label{sec:renorm}

In regions where $p(r)$ is constant, we have a flat spacetime
(although possibly with a deficit angle), and so to obtain
renormalized quantities we must only subtract the contribution of the
free Green's function.  As in the case of the point string, however,
the necessary cancellation may not appear term by term in the sum over
$\ell$, making numerical calculations difficult.

For $r>r_0$, we again use the approach of Ref.\
\cite{PhysRevD.42.2669} and consider the difference between the full
string and a point string with the same $\sigma$.  We can then add the
difference between the point string and empty space using Eq.\
(\ref{eqn:deltapointstring}).  We obtain, in both models,
\begin{equation}
G_\sigma(\bm{r},\bm{r}',\kappa)
-G_\sigma^{\rm point}(\bm{r},\bm{r}',\kappa)
= \frac{\sigma}{\pi} \zint \sum_{\ell=0}^\infty{}' B_{\wavevars,\ell}
K_{\sigma \ell}\left(\wavenumber r\right)
K_{\sigma \ell}\left(\wavenumber r'\right)
\cos \left[\ell (\theta-\theta')\right]
\end{equation}
for $r,r'>r_0$, written in terms of the scattering coefficient described
above for each model.  For the flowerpot model with $r<r_0$, we also
have flat space, although now corresponding to zero interior deficit angle
$\tilde \sigma = p(r) \sigma = 1$, and with the physical distance to the
origin given by $\displaystyle r_*=\frac{r}{\sigma}$.  We can
therefore subtract the free Green's function directly, evaluating it at
the same physical distances,
\begin{equation}
G_\sigma^{\rm flower}(\bm{r},\bm{r}',\kappa)
-G^{\rm free}(\bm{r}_*,\bm{r}'_*,\kappa)
= \frac{1}{\pi} \zint \sum_{\ell=0}^\infty{}' 
\frac{C_{\wavevars,\ell}^{\rm flower}}
{D_{\wavevars,\ell}^{\rm flower}}
I_{\sigma \ell}\left(\wavenumber r\right)
I_{\sigma \ell}\left(\wavenumber r'\right)
\cos \left[\ell (\theta-\theta')\right] \,.
\end{equation}
We can evaluate these expressions at coincident points, since
the singularity cancels through the subtraction.

For $r<r_0$ in the ballpoint pen model, we will use a hybrid of these 
subtractions.  First, we define
\begin{equation}
r_* = \frac{r_0}{\sqrt{\sigma^2 - 1}} \arccos \frac{1}{\sigma p(r)}
\qquad \Leftrightarrow \qquad
r = \frac{r_0 \sigma}{\sqrt{\sigma^2 - 1}} 
\sin \frac{r_* \sqrt{\sigma^2 - 1}}{r_0}
\label{eqn:rstar}
\end{equation}
so that $\displaystyle \frac{dr_*}{dr} = p(r)$
and $r_*$ represents the physical distance to the origin.  We then
subtract the contribution from a point string with deficit angle
$\tilde \sigma = \sigma p(r)$, corresponding to the angle deficit at
that point, evaluated at $r_*$.  As above, we add back in the
contribution of this point string using the results of the previous
section.

There is one further subtlety in this calculation.  The free Green's
function, what we ultimately subtract, depends only on the separation
between points, and thus is unchanged by translation or rescaling.
However, to carry out the subtraction, we must separate the points by
a distance $\epsilon$ in both Green's functions, and then take the
limit of the difference as $\epsilon$ goes to zero.  The limit should
correspond to splitting the points by the same physical distance.
Since
\begin{equation}
\lim_{\epsilon\to 0} \left[K_0(a\epsilon) - K_0(\epsilon)\right] =
- \log a \,,
\end{equation}
we therefore must subtract $\displaystyle \frac{1}{2\pi} 
\log \frac{r_*}{r p(r)}$ to correct for this discrepancy.

Thus we obtain, for $r<r_0$,
\begin{eqnarray}
G_\sigma^{\rm pen}(r,r,\kappa)
-G_{\tilde \sigma}^{\rm point}(r_*,r_*,\kappa) &\Rightarrow&
\frac{1}{\pi}\sum_{\ell=0}^\infty{}' \Bigg[
\frac{\Gamma(\nu(\wavevars)-\ell +1)}
{\Gamma(\nu(\wavevars)+\ell +1)}
P_{\nu(\wavevars)}^\ell\left(\tfrac{1}{\sigma  p(r)}\right)
\left(\frac{C_{\wavevars,\ell}^{\rm pen}}{D_{\wavevars,\ell}^{\rm pen}}
P_{\nu(\wavevars)}^\ell\left(\tfrac{1}{\sigma  p(r)}\right)
+ Q_{\nu(\wavevars)}^\ell\left(\tfrac{1}{\sigma  p(r)}\right)\right) 
\cr && - \tilde \sigma
I_{\tilde \sigma \ell}\left(\wavenumber r_*\right)
K_{\tilde \sigma \ell}\left(\wavenumber r_*\right)\Bigg]
-\frac{1}{2\pi}\log \frac{r_*}{r p(r)}
\end{eqnarray}
in the limit of coincident points.

\section{Renormalization: Tadpole Subtraction}
\label{sec:tadpole}

In the case of the ballpoint pen for $r<r_0$, the string background
effectively creates a background potential, leading to additional
counterterms.  Following Ref.\ \cite{OLUM2003175}, we use dimensional
regularization and consider configurations that are trivial in $n$
dimensions and spherically symmetric in $m$ dimensions, meaning that
a string in three space dimensions corresponds to the case
of $n=1$, $m=2$.  After integrating over the $n$ trivial directions, 
the contribution ${\cal G}_\ell(r,r',\kappa)$ to the Green's
function from angular momentum channel $\ell$ in $m$ dimensions
is replaced by the subtracted quantity
\begin{equation}
{\cal G}_{\ell,m}(r,r,\kappa) - {\cal G}^{\rm free}_{\ell,m}(r,r,\kappa)
+ (2-m)\frac{V_\ell(r)}{2\kappa^2} {\cal G}^{\rm free}_{\ell,m}(r,r,\kappa)
\end{equation}
where $V_\ell(r)$ is the background potential for that channel.  Here
the first subtraction represents the free background and the second
represents the tadpole graph.  Since we are interested in $m=2$, the latter
contribution appears to vanish. However, it multiplies the free
Green's function at coincident points, which diverges, so we must take
the limit carefully.  To do so, we consider the free radial Green's
function in $m$ dimensions for channel $\ell$,
\begin{equation}
{\cal G}^{\rm free}_{\ell,m}(r,r',\kappa) = 
\frac{\Gamma\left(\frac{m}{2}\right)}{2 \pi^{\frac{m}{2}}}
\frac{1}{(r r')^{\frac{m}{2}-1}} 
I_{\frac{m}{2} -1+\ell}(\kappa r_<) K_{\frac{m}{2} -1+\ell}(\kappa r_>)\,.
\end{equation}
Its contribution is weighted by the degeneracy factor
\begin{equation}
D^m_\ell = \frac{\Gamma(m+\ell-2)}{\Gamma(m-1)\Gamma(\ell+1)}(m+2\ell-2) \,,
\end{equation}
which has the following limits as special cases 
\begin{equation}
D^{m=2}_\ell = 2 - \delta_{\ell 0} \qquad
D^{m=1}_\ell = \delta_{\ell 0} + \delta_{\ell 1} \qquad
D^{m=0}_\ell = \delta_{\ell 0}\,,
\end{equation}
expressed in terms of the Kronecker $\delta$ symbol.  For equal angles,
the free Green's function is then given by the sum
\begin{equation}
G^{\rm free}_{m}(r,r',\kappa) = \frac{1}{(2\pi)^{\frac{m}{2}}}
\left(\frac{\kappa}{|r-r'|}\right)^{\frac{m}{2} - 1} 
K_{\frac{m}{2} -1}(\kappa |r-r'|) 
= \sum_{\ell=0}^\infty D^m_\ell
{\cal G}^{\rm free}_{\ell,m}(r,r',\kappa)\,.
\end{equation}
To bring the points together, we expand around $r=r'=0$ (since the
free Green's function only depends on their difference, we may choose 
to have them both approach any point we choose), in which case
we have
\begin{equation}
(2-m) D^m_\ell
{\cal G}^{\rm free}_{\ell,m}(r,r,\kappa) = -(2-m)\left[
\left(\frac{\kappa r}{2}\right)^{2\ell}
\frac{\kappa ^{m-2}
\left(m+2 \ell\right)  \Gamma\left(\frac{m}{2}\right)
\Gamma\left(2-\frac{m}{2}-\ell \right) \Gamma(m + \ell - 2)}
{(4\pi)^{\frac{m}{2}} 
\Gamma(m-1) \Gamma(\ell+1) \Gamma\left(1+\frac{m}{2}+\ell \right)}
+ {\cal O}(r^{2-m})\right]\,,
\end{equation}
where, crucially, we have dropped terms of order $r^{2-m}$ because we
approach $m=2$ from below, where the integrals converge, and so these
terms vanish for $r\to 0$.  When $\ell \neq 0$, the term we have kept also
vanishes for $r\to 0$. However, for $\ell=0$, it goes to
$\displaystyle \frac{1}{2\pi}$ in the limit $m\to 2$, and so we have
found
\begin{equation}
\lim_{m\to 2} \left[(2-m) D^m_\ell
 {\cal G}^{\rm free}_{\ell,m}(r=0,r=0,\kappa) \right] =
\frac{1}{2\pi} \delta_{\ell 0}
\quad \Longrightarrow \quad
\lim_{m\to 2}
\left[(2-m) \sum_{\ell=0}^\infty D^m_\ell
{\cal G}^{\rm free}_{\ell,m}(r,r,\kappa) V_\ell(r)\right] =
\frac{1}{2\pi}V_{\ell=0}(r)\,,
\end{equation}
with the result for the summed Green's function depending only on the 
contribution of the potential in the $\ell=0$ channel.

To find the potential $V_\ell(r)$, we rewrite the wave equation for
$r<r_0$ using the physical radius $r_*$ given in Eq.\
(\ref{eqn:rstar}). The rescaled wavefunction
$\displaystyle \phi_{\wavevars,\ell}(r_*) =
 \sqrt{r} \psi_{\wavevars,\ell}(r)$ then obeys
\cite{PhysRevD.59.064017,Khusnutdinov:2004ux}
\begin{equation}
\left[-\frac{d^2}{dr_*^2} + \sigma^2
\left(\frac{\ell^2-\frac{1}{4}}{r^2}\right) 
-\frac{1}{4} \left(\frac{\sigma^2-1}{r_0^2}\right)
+\xi {\cal R} + \wavenumbersq \right] \phi_{\wavevars,\ell}(r_*) = 0\,,
\end{equation}
where the denominator of the second term represents $r$ as a function
of $r_*$.  A free particle would instead obey the equation
\begin{equation}
\left[-\frac{d^2}{dr_*^2} + \left(\frac{\ell^2-\frac{1}{4}}{r_*^2}\right) 
+ \wavenumbersq \right] \phi^{\rm free}_{\wavevars,\ell}(r_*) = 0\,,
\end{equation}
and so we can consider the difference between the two expressions in
brackets as a scattering potential,
\begin{equation}
V_\ell^{\rm full}(r) = \frac{\left(\ell^2-\frac{1}{4}\right) \sigma ^2}{r^2}
-\frac{\left(\ell^2-\frac{1}{4}\right) 
\left(\sigma^2-1\right)}{r_0^2 
\left(\arccos\frac{1}{\sigma p(r)}\right)^2}
+ \frac{(8 \xi - 1) \left(\sigma ^2 - 1\right)}{4 r_0^2} \,.
\end{equation}
For the counterterm, we need only the  $\ell=0$ case, and the tadpole
subtraction is given by the leading order in perturbation
theory.  We take $\sigma^2-1$ as the coupling constant. Expanding to
leading order in this quantity, we obtain the tadpole contribution for
$r<r_0$,
\begin{equation}
V_{\ell=0}(r) = 2\frac{\sigma^2 - 1}{r_0^2} \left(\xi -\frac{1}{6}\right) 
= {\cal R} \left(\xi - \frac{1}{6}\right) \,,
\end{equation}
which is independent of $r$ and vanishes for conformal coupling in
three dimensions.  Note that this term exactly coincides with the
first-order heat kernel coefficient \cite{BrlDv}.

Putting these results together, we obtain the subtracted
Green's function summed over angular momentum channels
\begin{equation}
G_\sigma(r,r,\kappa) -
G^{\rm free}(r,r,\kappa) + \frac{V_{\ell=0}(r)}{4\pi \kappa^2} 
= G_\sigma(r,r,\kappa) - G^{\rm free}(r,r,\kappa)
+ \frac{{\cal R}}{4\pi \kappa^2} \left(\xi - \frac{1}{6}\right)
\label{eqn:greensub}
\end{equation}
in the limit where $m\to 2$ and the points are coincident.
Furthermore, we can pull the last term of Eq.\ (\ref{eqn:greensub})
inside the sum used to define the Green's function by using a special
case of the addition theorem for Legendre functions,
\begin{equation}
1 = 2 \sum_{\ell=0}^\infty{}'
\frac{\Gamma(\nu(\wavevars)-\ell +1)} {\Gamma(\nu(\wavevars)+\ell +1)}
P_{\nu(\wavevars)}^\ell\left(\tfrac{1}{\sigma  p(r)}\right)^2 \,.
\label{eqn:legendreidentity}
\end{equation}

\section{Derivative Term}
\label{sec:deriv}

Next we compute the derivative term $\displaystyle
\frac{1}{r^2} {\cal D}_r^2 G_\sigma(\bm{r},\bm{r}, \kappa)$ by
differentiating the expressions above.  We note that for any two
functions $\psi^A(r)$ and $\psi^B(r)$,
\begin{equation}
{\cal D}_r^2 (\psi^A(r) \psi^B(r)) = 
2({\cal D}_r \psi^A(r))({\cal D}_r \psi^B(r)) + 
\psi^A(r) \left({\cal D}_r^2 \psi^B(r)\right) + 
\left({\cal D}_r^2 \psi^A(r) \right) \psi^B(r) \,,
\end{equation}
and so by using the equations of motion, we can express the terms
involving squares of first derivatives in terms of the second
derivative of the product, and vice versa.  Accordingly, for any pair
of solutions $\psi_{\wavevars,\ell}^{A}(r)$ and
$\psi_{\wavevars,\ell}^{B}(r)$ obeying Eq.\ (\ref{eqn:eom}), we have
\begin{equation}
\frac{1}{2 r^2} {\cal D}_r^2
\left(\psi_{\wavevars,\ell}^{A} (r)
\psi_{\wavevars,\ell}^{B}(r)\right)
= \left(\frac{(\sigma \ell)^2}{r^2} + \xi {\cal R} + \wavenumbersq\right)
\psi_{\wavevars,\ell}^{A}(r) \psi_{\wavevars,\ell}^{B}(r)
+ \frac{1}{r^2}\left({\cal D}_r 
\psi_{\wavevars,\ell}^{A}(r)\right)
\left({\cal D}_r\psi_{\wavevars,\ell}^{B}(r)\right) \,,
\end{equation}
and we note that
\begin{equation}
\frac{1}{p(r)}\frac{d}{dr} 
P_{\nu(\wavevars)}^\ell\left(\tfrac{1}{\sigma  p(r)}\right) =
-\frac{r(\sigma^2-1)}{r_0^2 \sigma}
P_{\nu(\wavevars)}^\ell{}'\left(\tfrac{1}{\sigma  p(r)}\right)\,,
\end{equation}
and similarly for $Q_{\nu(\wavevars)}^\ell$, and so by using
recurrence relations we can simplify
\begin{eqnarray}
\frac{1}{2 r^2} {\cal D}_r^2 \left[
P_{\nu(\wavevars)}^\ell\left(\tfrac{1}{\sigma  p(r)}\right)
Z_{\nu(\wavevars)}^\ell\left(\tfrac{1}{\sigma  p(r)}\right)\right] &=&
\left(\frac{(\sigma \ell)^2}{r^2} + \xi {\cal R} + \wavenumbersq\right)
P_{\nu(\wavevars)}^\ell\left(\tfrac{1}{\sigma  p(r)}\right)
Z_{\nu(\wavevars)}^\ell\left(\tfrac{1}{\sigma  p(r)}\right) \cr
&&+\frac{1}{r^2} 
\left[\frac{r\sqrt{\sigma^2-1}}{r_0}
P_{\nu(\wavevars)}^{\ell+1}\left(\tfrac{1}{\sigma  p(r)}\right)
+ \frac{\ell}{p(r)}
P_{\nu(\wavevars)}^\ell\left(\tfrac{1}{\sigma  p(r)}\right)\right] \cr
&& \times \left[\frac{r\sqrt{\sigma^2-1}}{r_0}
Z_{\nu(\wavevars)}^{\ell+1}\left(\tfrac{1}{\sigma  p(r)}\right)
+ \frac{\ell}{p(r)}
Z_{\nu(\wavevars)}^\ell\left(\tfrac{1}{\sigma  p(r)}\right) \right] \,,
\end{eqnarray}
where $Z$ is either $P$ or $Q$; similar simplifications based on
recurrence relations apply for Bessel functions.

Renormalization of the derivative term in the curved space background
requires an additional counterterm compared to the flat space
expression in Eq.\ (\ref{eqn:onesub}).  This subtraction is also
proportional to the curvature scalar ${\cal R}$.  The renormalized derivative
term becomes 
\begin{equation}
\frac{1}{r^2}{\cal D}_r^2 G_\sigma(r,r,\kappa)
 - \frac{1}{4\pi} {\cal R} \,,
\label{eqn:derivcounterterm}
\end{equation}
where again the counterterm is proportional to the Ricci scalar and we
have taken the points to be coincident.  With this choice, the full
tadpole counterterm contribution to the modified integrand of Eq.\
(\ref{eqn:onesub}) with $m=2$ and $n=0$ is
\begin{equation}
\kappa^2\frac{{\cal R}}{4\pi \kappa^2} \left(\xi - \frac{1}{6}\right) +
\left(\frac{1}{4}-\xi\right) \frac{{\cal R}}{4\pi}
=\frac{{\cal R}}{48\pi} \,,
\label{eqn:counterterm}
\end{equation}
consistent with the general result originating from the
two-dimensional conformal anomaly
\cite{PhysRevD.13.2720,1977RSPSA.354...59D,PhysRevD.15.2088}, since
our geometry is only curved in two dimensions.\footnote{For
a general entry in the stress energy tensor 
$T_{\alpha\beta}$, this counterterm would become $\displaystyle 
\frac{\xi}{2\pi} R_{\alpha\beta}- \frac{{\cal R}}{48\pi}g_{\alpha\beta}$.}
As above, we can pull this
term inside the sum using Eq.\ (\ref{eqn:legendreidentity}).

Finally, as before we find it more computationally tractable to
compute the derivative term for the difference of the full
Green's function and the corresponding point string.  We must then add
back the derivative of the point string contribution as well.  In both
what we subtract and add back in, we define the radial derivative
for the point string contribution taking $p(r)$ constant,
corresponding to the derivative we would use in the point string case.

\section{Kontorovich-Lebedev Approach for Nonzero Width String}
\label{sec:KL}

As with the point string above, we can express the Green's function as
an integral over imaginary angular momentum $\lambda$ using the
Kontorovich-Lebedev approach.  As described above, we take the pairs
$P_{\nu(\wavevars)}^\ell$ and $P_{\nu(\wavevars)}^{-\ell}$ and
$I_{\ell}$ and $I_{-\ell}$ 
as the independent solutions for $r<r_0$ in the ballpoint pen and
flowerpot models respectively.

The Green's function becomes
\begin{equation}
G_\sigma(\bm{r},\bm{r}',\kappa) = \frac{1}{2\pi} \zint \int_0^\infty 
\frac{id\lambda}{\sinh \frac{\lambda \pi}{\sigma}}
\left[\psi_{\wavevars,i\frac{\lambda}{\sigma}}^{\rm reg} (r_<) -
\psi_{\wavevars,-i\frac{\lambda}{\sigma}}^{\rm reg} (r_<) \right]
\psi_{\wavevars,i\frac{\lambda}{\sigma}}^{\rm out} (r_>)
\cosh \left[\lambda\left(\frac{\pi}{\sigma} -
|\theta-\theta'|\right)\right] \,,
\end{equation}
where we have used that the outgoing wave is always even in $\lambda$.
For $r>r_0$, we can use Wronskian relationships to simplify
\begin{equation}
\frac{i}{\sinh \frac{\lambda \pi}{\sigma}}
\left[\psi_{\wavevars,i\frac{\lambda}{\sigma}}^{\rm reg,pen} (r) -
\psi_{\wavevars,-i\frac{\lambda}{\sigma}}^{\rm reg,pen} (r) \right] =
\frac{\frac{2}{\pi} \sigma^3 K_{i\lambda}\left(\wavenumber r\right)}
{|(\sigma^2-1)P_{\nu(\wavevars)}^{i\frac{\lambda}{\sigma}}
{}'\left(\tfrac{1}{\sigma}\right)
K_{i\lambda}\left(\wavenumber r_0\right) +
\sigma \kappa r_0
P_{\nu(\wavevars)}^{i\frac{\lambda}{\sigma}}\left(\tfrac{1}{\sigma}\right)
K_{i\lambda}'\left(\wavenumber r_0\right)|^2}
\end{equation}
for the ballpoint pen and
\begin{equation}
\frac{i}{\sinh \frac{\lambda \pi}{\sigma}}
\left[\psi_{\wavevars,i\frac{\lambda}{\sigma}}^{\rm reg,flower} (r) -
\psi_{\wavevars,-i\frac{\lambda}{\sigma}}^{\rm reg,flower} (r) \right] =
\frac{\frac{2}{\pi} \frac{\sigma}{\kappa^2 r_0^2} 
K_{i\lambda}\left(\wavenumber r\right)}
{|I_{i\frac{\lambda}{\sigma}}'\left(\wavenumber \frac{r_0}{\sigma}\right)
K_{i\lambda}\left(\wavenumber r_0\right) 
- I_{i\frac{\lambda}{\sigma}}\left(\wavenumber \frac{r_0}{\sigma}\right)
K_{i\lambda}'\left(\wavenumber r_0\right)
+ \frac{2 \xi(\sigma-1)}{\wavenumber r_0}
I_{i\frac{\lambda}{\sigma}}\left(\wavenumber \frac{r_0}{\sigma}\right)
K_{i\lambda}\left(\wavenumber r_0\right)|^2}
\end{equation}
for the flowerpot.  For $r<r_0$, we have
\begin{equation}
\frac{i}{\sinh \frac{\lambda \pi}{\sigma}}
\left[\psi_{\wavevars,i\frac{\lambda}{\sigma}}^{\rm reg} (r) -
\psi_{\wavevars,-i\frac{\lambda}{\sigma}}^{\rm reg} (r) \right] =
\frac{i}{\sinh \frac{\lambda \pi}{\sigma}}
\frac{A_{\wavevars,\frac{i\lambda}{\sigma}}}
{C_{\wavevars,\frac{i\lambda}{\sigma}}}
\psi_{\wavevars,i\frac{\lambda}{\sigma}}^{\rm out} (r) \,,
\end{equation}
and so in both cases the Green's function is written entirely in terms
of outgoing waves.  We can then subtract the free Green's function in
the form of Eq.~(\ref{eqn:KLfree}).  For numerical calculation,
however, we find that this approach is only effective for $r>r_0$.

\section{Results}

Collecting all of these terms, we have the full expression for the
renormalized energy density, written with the point string subtracted
and then added back in,
\begin{eqnarray}
\langle{\cal H}\rangle_{\rm ren} &=& -\frac{1}{\pi} \int_0^\infty d\kappa
\Bigg[\kappa^2 \left( G_\sigma(r,r,\kappa)
-G_{\tilde \sigma}^{\rm point}(r_*,r_*,\kappa)
+\Delta G_{\tilde \sigma}^{\rm point}(r_*,r_*,\kappa)
-\frac{1}{2\pi}\log \frac{r_*}{r p(r)} \right) \cr
&&-\left(\frac{1}{4}-\xi\right) \frac{1}{r^2}\left( 
{\cal D}_r^2 G_\sigma(r,r,\kappa)
-\bar{\cal D}_r^2 G_{\tilde \sigma}^{\rm point}(r_*,r_*,\kappa)
+\bar{\cal D}_r^2 \Delta G_{\tilde \sigma}^{\rm point}(r_*,r_*,\kappa)
\right) + \frac{{\cal R}}{48\pi}\Bigg] \,,
\label{eqn:etotal}
\end{eqnarray}
where $r_*$ is the physical distance in each model as defined above
(with $r_*=r$ for $r>r_0$) and $\tilde \sigma = p(r) \sigma$ in each
region (with $\tilde \sigma =\sigma$ for $r>r_0$).  Here we have
defined $\displaystyle \bar{\cal D}_r = r \frac{d}{dr}$ so that we add
and subtract derivatives of the point string in the background of a flat
spacetime with a deficit angle, as described above,
and we denote $\displaystyle \bar{\cal D}_r^2 G_{\tilde \sigma}^{\rm point}
(r_*, r_*, \kappa) = \bar{\cal D}_r^2 G_{\tilde \sigma}^{\rm point}
(r, r,\kappa)\vert_{r=r_*}$, and similarly for other derivatives at $r_*$.
The combined counterterm $\displaystyle \frac{{\cal R}}{48\pi}$ 
(with ${\cal R}=0$ for $r>r_0$, and for $r<r_0$ in the flowerpot
model) is obtained by combining the two individual terms obtained in
Sec.\ \ref{sec:tadpole} using Eq.\ (\ref{eqn:counterterm}).

In both the first and second lines of Eq.\ (\ref{eqn:etotal}),
the contribution from the difference between the full and point string
Green's functions can be taken inside the sum over
$\ell$, using the results in Sec.\ \ref{sec:renorm} and
Eq.\ (\ref{eqn:pointstring}), while the contribution from
the difference between the point string and empty space Green's
functions can be computed as an integral over imaginary angular momentum
using Eq.\ (\ref{eqn:deltapointstring}).  For the case of $r>r_0$, we
can check our calculation using the results of Sec.\ \ref{sec:KL}, in
which case Eq.\ (\ref{eqn:etotal}) can be expressed entirely in terms of an
integral on the imaginary angular momentum axis.  For that
calculation, there is no need to add and subtract the point string
contribution, so we can simply subtract the free Green's function
directly, using Eq.\ (\ref{eqn:KLfree}).

\begin{figure}[htbp]
\includegraphics[width=0.45\linewidth]{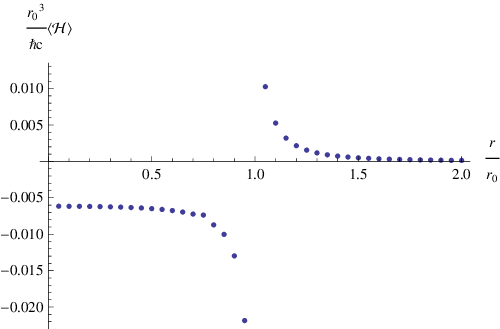} \hfill
\includegraphics[width=0.45\linewidth]{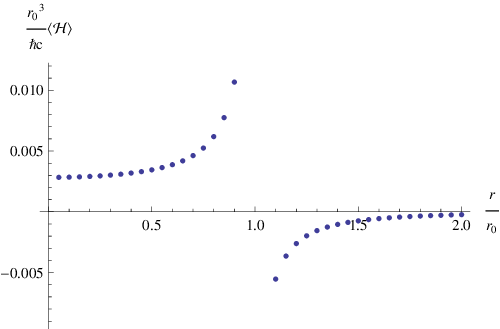}
\caption{Energy density $\langle \cal H\rangle_{\rm ren}$, in units of
$\displaystyle \frac{\hbar c}{r_0^3}$, as a function of $r$, in units of
$r_0$, for $\displaystyle \theta_0 = \frac{\pi}{3}$ in the ballpoint
pen model.  The left panel shows minimal coupling $\xi=0$, while the
right panel shows conformal coupling $\displaystyle \xi = \frac{1}{8}$.}
\label{fig:energyp2}
\end{figure}

\begin{figure}[htbp]
\includegraphics[width=0.45\linewidth]{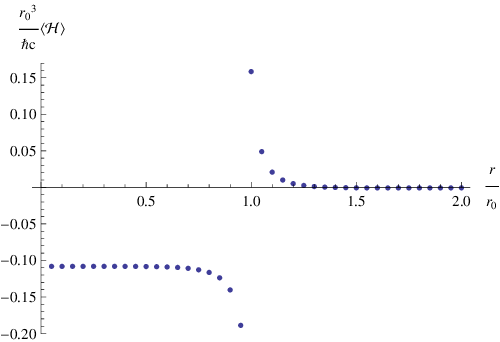} \hfill
\includegraphics[width=0.45\linewidth]{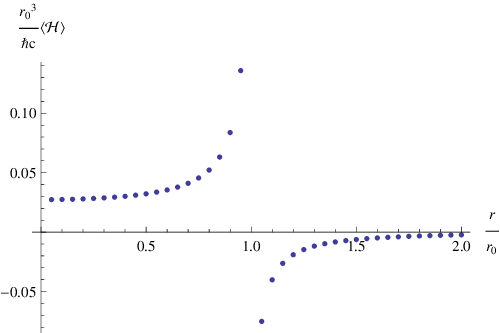}
\caption{Energy density $\langle \cal H\rangle_{\rm ren}$, in units of
$\displaystyle \frac{\hbar c}{r_0^3}$, as a function of $r$, in units of
$r_0$, for $\displaystyle \theta_0 = \frac{2\pi}{3}$ in the ballpoint
pen model.  The left panel shows minimal coupling $\xi=0$, while the
right panel shows conformal coupling $\displaystyle \xi =
\frac{1}{8}$.  The energy shows a similar shape, but larger magnitude
for a greater deficit angle.}
\label{fig:energyp3}
\end{figure}

\begin{figure}[htbp]
\includegraphics[width=0.45\linewidth]{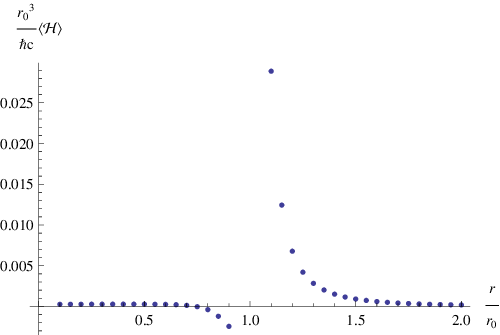} \hfill
\includegraphics[width=0.45\linewidth]{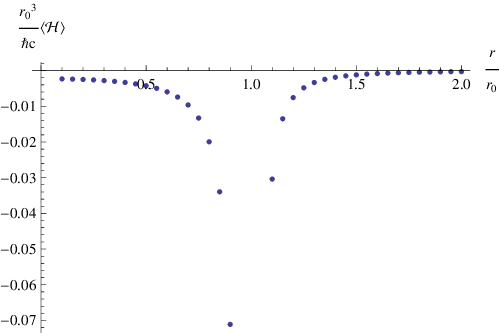}
\caption{Energy density $\langle \cal H\rangle_{\rm ren}$, in units of
$\displaystyle \frac{\hbar c}{r_0^3}$, as a function of $r$, in units of
$r_0$, for $\displaystyle \theta_0 = \frac{\pi}{3}$ in the flowerpot
model.  The left panel shows minimal coupling $\xi=0$, while the
right panel shows conformal coupling $\displaystyle \xi =
\frac{1}{8}$.  The energy density for minimal coupling is
small for $r<r_0$ because this case is close to the deficit angle
where the inside energy density changes sign.}
\label{fig:energyf2}
\end{figure}

\begin{figure}[htbp]
\includegraphics[width=0.45\linewidth]{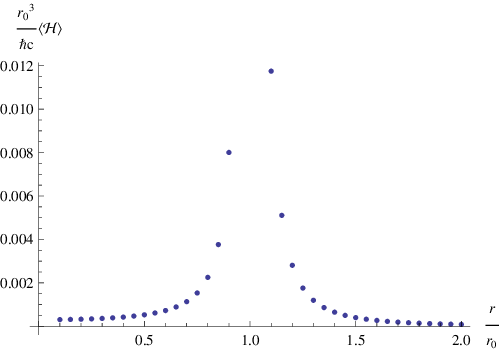} \hfill
\includegraphics[width=0.45\linewidth]{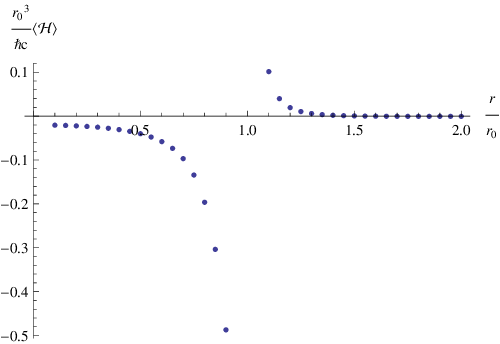}
\caption{Energy density $\langle \cal H\rangle_{\rm ren}$, in units of
$\displaystyle \frac{\hbar c}{r_0^3}$, as a function of $r$, in units of
$r_0$, for $\displaystyle \theta_0 = \frac{\pi}{6}$ (left panel) and
$\displaystyle \theta_0 = \frac{2\pi}{3}$ (right panel
in the flowerpot model with $\xi=0$.  The sign of the energy density
for $r<r_0$ reverses at approximately $\theta_0 \approx 1$.}
\label{fig:energyfsign}
\end{figure}

Sample results are shown in Figs.\ \ref{fig:energyp2} through
\ref{fig:energyfsign}, for both minimal and conformal coupling.  We note
that for the interior of the flowerpot, the sign of the energy density
for $r<r_0$ is opposite at large and small deficit angles for minimal
coupling, with the sign change occurring at $\theta_0 \approx 1$.  For
the ballpoint pen, we see a singularity at $r=r_0$, corresponding to
the step function discontinuity in the curvature.  However, the total
energy remains finite, because this contribution cancels, as a
principal value, on either side of the boundary \cite{OLUM2003175}.
In an actual string, the sharp edge would be smoothed by both the
classical string dynamics and the back-reaction from the quantum
field.  For the flowerpot, the energy density diverges at $r=r_0$ and
the total energy is divergent as well, because the curvature profile
is itself a divergent $\delta$-function, which gives rise to an
infinite quantum total energy \cite{density}.  As a result, in this
case the energy density need not have opposite signs for $r<r_0$ and
$r>r_0$.  It is interesting to note that the effects of the string's
curvature are qualitatively similar to those of the analogous square
well or $\delta$-function scalar background potential, as studied in
Refs.\ \cite{density,OLUM2003175}.

\section{Conclusions}
We have shown how to use scattering data to compute the quantum
energy density of a massless scalar field in the background on a 
nonzero width cosmic string background, using both the flowerpot
and ballpoint pen string profiles in two space dimensions.  Of
particular interest is the interior of the ballpoint pen, where 
the background space time has nontrivial (but constant) curvature.  We
precisely specify counterterms corresponding to renormalization of
both the cosmological constant and the gravitational coupling to the
scalar curvature ${\cal R}$.  In addition, to make the calculation tractable
numerically, we subtract and then add back in the contribution of a
point string with the same deficit angle and physical radius.
We can then subtract the free space contribution, corresponding to the
cosmological constant renormalization, by combining it with the point
string result and using analytic continuation of the angular momentum
sum to an integral over the imaginary axis. These results extend
straightforwardly to three dimensions, but that case requires an
additional subtraction of order ${\cal R}^2$.

\section{Acknowledgments}
It is a pleasure to thank K.\ Olum for sharing preliminary work on
this topic and H.\ Weigel for helpful conversations and feedback. 
M.\ K., X.\ L., and N.\ G.\ were supported in part by the National
Science Foundation (NSF) through grant PHY-2209582.

\bibliographystyle{apsrev}
\bibliography{flowerpen}

\end{document}